# INTENSITY CONSERVING SPECTRAL FITTING

J. A. Klimchuk
NASA Goddard Space Flight Center, Greenbelt, MD 20771, USA;
James.A.Klimchuk@nasa.gov

S. Patsourakos
Physics Department, University of Ioannina, Ioannina GR-45110, Greece

D. Tripathi
Inter-University Centre for Astronomy and Astrophysics, Post Bag-4, Ganeshkhind, Pune 411007, India

## ABSTRACT

The detailed shapes of spectral line profiles provide valuable information about the emitting plasma, especially when the plasma contains an unresolved mixture of velocities, temperatures, and densities. As a result of finite spectral resolution, the intensity measured by a spectrometer is the average intensity across a wavelength bin of non-zero size. It is assigned to the wavelength position at the center of the bin. However, the actual intensity at that discrete position will be different if the profile is curved, as it invariably is. Standard fitting routines (spline, Gaussian, etc.) do not account for this difference, and this can result in significant errors when making sensitive measurements. Detection of asymmetries in solar coronal emission lines is one example. Removal of line blends is another. We have developed an iterative procedure that corrects for this effect. It can be used with any fitting function, but we employ a cubic spline in a new analysis routine called Intensity Conserving Spline Interpolation (ICSI). As the name implies, it conserves the observed intensity within each wavelength bin, which ordinary fits do not. Given the rapid convergence, speed of computation, and ease of use, we suggest that ICSI be made a standard component of the processing pipeline for spectroscopic data.

## 1.  INTRODUCTION

Spectroscopy is an extremely powerful means of diagnosing astrophysical and other plasmas. It is especially valuable when the emitting plasma has a mixture of velocities, temperatures, and densities along an optically thin line-of-sight or across the finite area of a resolution element. This is of course the case for stellar observations, where the emission is integrated over the entire star, but it is also true for many solar observations, where a typical pixel represents an average over many elemental structures. Coronal loops are a prime example. Each loop is believed to be a bundle of unresolved magnetic strands that are heated by small bursts of energy called nanoflares (e.g., Klimchuk 2006, 2015; Reale 2010). The nanoflares produce a characteristic pattern of fast upflows and slow downflows, and therefore multiple out-of-phase events give rise to spectral lines with non-Gaussian, asymmetric profiles (Patsourakos & Klimchuk 2006). By studying the shapes of the profiles, we can hope to infer the distribution of unresolved flows that are present.

Considerable effort has been devoted recently to measuring the asymmetries of EUV line profiles from the solar corona and transition region. One goal of these investigations is to detect the hot tips of type II spicules---otherwise cool jets that are ejected rapidly upward from the chromosphere (de Pontieu et al. 2009). As in the nanoflare scenario, the hot plasma is expected to slowly drain back down to the surface. The combined emission from strongly blueshifted upflows and weakly redshifted downflows produces asymmetric line profiles. The degree of asymmetry is an important test of whether spicules provide significant amounts of hot plasma to the corona (Klimchuk 2012; Klimchuk & Bradshaw 2014; Bradshaw & Klimchuk 2015).

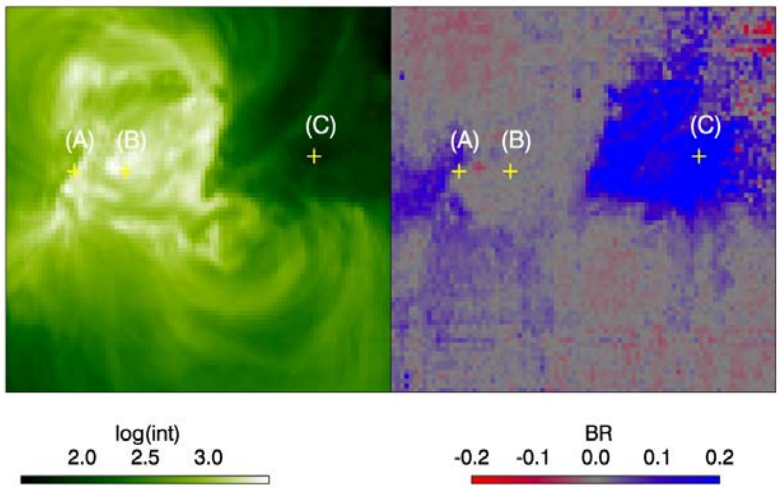

**Figure 1.** Intensity (left) and blue-red asymmetry (right) observed in Fe XIV (274 A) for active region NOAA AR 10978 observed on 2007 December 11. Lettered crosses indicate the locations of line profiles in Figures 3, 5, and 6.

A parameter called the blue-red (BR) asymmetry is typically used to quantify the measurements. It is defined to be the difference in the emission integrated over wavelength (velocity) intervals in the blue and red wings, divided by the emission integrated over the line core. The specific intervals vary from study to study, but a common wing interval is $\pm[50, 150]$ km s$^{-1}$, and a common core interval is $[-30, 30]$ km s$^{-1}$. Standard convention is that positive asymmetries mean excess emission in the blue wing. The amplitudes of the reported asymmetries are generally very small, < 5%, but in some faint areas at the perimeters of active regions they can exceed 20% (Hara et al. 2008; De Pontieu et al. 2009; McIntosh & De Pontieu 2009; Bryans et al. 2010; De Pontieu et al. 2011; Tian et al. 2011; Martinez-Sykora et al. 2011; Doschek 2012; Tian et al. 2012; Brooks & Warren 2012; Tripathi & Klimchuk 2013; Patsourakos et al. 2014). Figure 1 shows intensities and asymmetries measured in the Fe XIV (274 A) line, formed at ~ 2 MK, using 256x256 arcsec$^2$ raster data from the Extreme-ultraviolet Imaging Spectrometer (EIS) on the Hinode spacecraft (Culhane et al. 2007). This is the same active region studied by Brooks & Warren (2012), Patsourakos et al. (2014), and Kitagawa & Yokoyama (2015).

A major challenge in these asymmetry studies concerns the choice of the velocity zero point. Most space-based observations do not have an absolute wavelength calibration. Moreover, the velocity scale often drifts during the course of a spacecraft orbit due to variable thermal stresses experienced by the instrument. A common solution is to define the rest wavelength, $\lambda_0$, to be the location of peak intensity in the line profile. Since the measured asymmetries are typically small, $\lambda_0$ must be determined with a high accuracy---much better than the wavelength resolution. One approach is to perform a Gaussian fit. This has obvious dangers, since the fit is affected by the asymmetry that is being measured. Restricting the fit to the line core can help, but problems remain unless the asymmetry is known to be present only in outer wings of the profile (Tripathi & Klimchuk 2013). Double Gaussian fits offer an improvement, but they are still only an approximation to the real shape of the profile. A better approach for finding the position of peak intensity is to perform a spline interpolation. It makes the least restrictive assumptions about the profile shape. Spline interpolation is also useful for performing the wing and line core intensity integrations, since velocity intervals do not coincide with wavelength grid of the data.

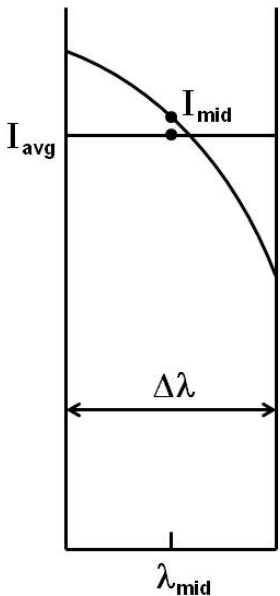

**Figure 2.** Schematic drawing of a portion of a spectral line profile contained within a single wavelength bin.

Over the course of measuring BR asymmetries in solar active regions (Tripathi & Klimchuk 2013; Patsourakos et al. 2014), we came to realize a fundamental misconception in the way spectral data are interpreted and used. Consider the schematic drawing in Figure 2. It represents a section of a line profile corresponding to a single wavelength bin. The spectral resolution of the observations is $\Delta\lambda$. If the profile has downward concavity within the bin, as shown, its intensity $I_{mid}$ at the bin midpoint $\lambda_{mid}$ is greater than the average intensity $I_{avg}$. The quantity measured by the spectrometer is $I_{avg}$, since the detector integrates the signal over the unresolved interval $\Delta\lambda$. Problems arise because this average intensity is traditionally assigned to the midpoint wavelength position. To be accurate, either $I_{avg}$ should be assigned to a longer wavelength (where the profile in Fig. 2 intersects the horizontal line at $I_{avg}$), or the intensity assigned to the midpoint should be revised upward from $I_{avg}$.

We have devised a simple iterative procedure for making the latter correction. We call it Intensity Conserving Spline Interpolation (ICSI). As the name implies, ICSI conserves the intensity of the original data. The spline fit obtained with ICSI has an average over $\Delta\lambda$ that is equal to the observed (true) value $I_{avg}$. A traditional spline fit passes through $I_{avg}$ at $\lambda_{mid}$ and therefore does not conserve the original intensity. Such a fit would lie below the true profile in Figure

2. It would lie above the true profile if the profile were concave upward. We choose a cubic spline because it provides the smoothest fit (minimizes the integral of the second derivative), though any order spine, or indeed any fitting function, can be used.

Errors occur when the effects corrected by ICSI are not taken into account. In general, they are small, but they can be very significant when highly accurate measurements are required. For example, the errors in BR asymmetry can be comparable to the asymmetries themselves. The determination of components in multi-component profiles (e.g., blended lines) is another example. Both are discussed below after a description of the ICSI procedure is presented.

## 2. ICSI PROCEDURE

As discussed above, the intensity observed by a spectrometer is the average intensity over the wavelength bin $\Delta\lambda$:

$$I_{obs} = I_{avg} = \frac{1}{\Delta\lambda}\int_{\Delta\lambda} I\, d\lambda . \qquad (1)$$

It is understood that there is a different intensity for each bin, i.e., $I_{obs} = I_{obs}(\lambda_j)$ at bin $j$. Let $F^0$ be the traditional spline fit of the data, which passes directly through every point in the profile. The intensity of the fit at the midpoint of the wavelength bin is $F_{mid}^{0} = I_{obs}$, and the bin-averaged intensity is

$$F_{avg}^{0} = \frac{1}{\Delta\lambda}\int_{\Delta\lambda} F^0\, d\lambda . \qquad (2)$$

As explained above, $F_{avg}^{0}$ will be different from $I_{obs}$, yet we want an eventual solution where the two are equal. We therefore correct the midpoint intensity according to

$$I_{cor} = I_{obs}\frac{F_{mid}^{0}}{F_{avg}^{0}} . \qquad (3)$$

We then use the corrected value to perform an improved spline fit $F^1$. The process can be repeated as many times as desired using the iterative formula

$$F_{mid}^{i+1} = I_{obs}\frac{F_{mid}^{i}}{F_{avg}^{i}} . \qquad (4)$$

The solution improves with each successive iteration. By rewriting the formula as

$$F_{avg}^i = I_{obs} \frac{F_{mid}^i}{F_{mid}^{i+1}}, \tag{5}$$

we see that, as $i \rightarrow \infty$, the solution converges such that $F_{mid}^{i+1} = F_{mid}^i$ and $F_{avg} = I_{obs}$. The final spline fit conserves the intensity of the original data to any desired accuracy. This is not the case for the first spline fit at the start of the iterative process.

We find that the solution converges quickly in all cases we have studied. Figure 3 shows a profile from raster position (71,140) in Figure 1 (position B). The stars are the data, the dotted curve is the traditional spline fit (without ICSI), and the solid curve is the fit after 6 iterations. Figure 4 shows the largest intensity error ($F_{avg}^i - I_{avg}$) across the profile, normalized by the maximum intensity, as a function of iteration number.[1] We consider the solution to be converged when the error drops below $10^{-5}$.

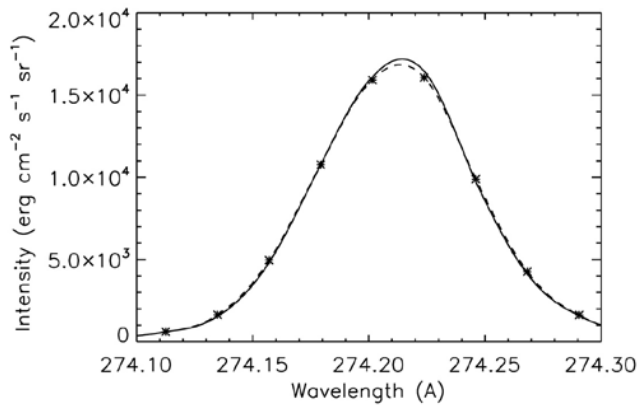

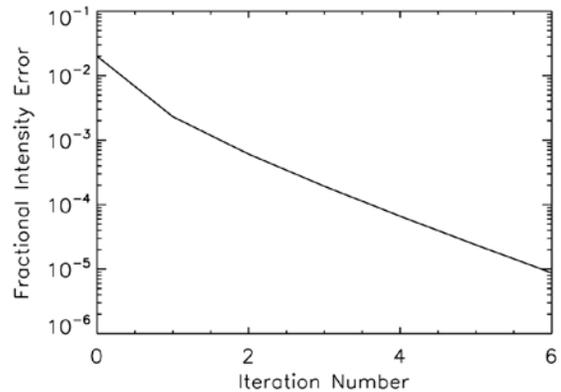

**Figure 3.** Line profile from raster position B in Figure 1. Stars--observed intensities; dashed curve-- traditional spline fit; solid curve--spline fit with ICSI.

**Figure 4.** Largest intensity error ($F_{avg}^i - I_{avg}$) across the profile, normalized by the maximum intensity, as a function of iteration number for the profile in Figure 3.

Note that the ICSI solution has the properties described in the Introduction. It lies above the data points and traditional fit where the profile is concave downward, and below the data points and traditional fit where the profile is concave upward. The greatest deviation is near the peak of the profile, where the curvature is greatest. Although the differences seem subtle, they have an important effect on the BR asymmetry. The ICSI solution has an asymmetry of 3.9%, whereas the traditional fit has an asymmetry of only 2.6%. Here, as in most cases, the error is driven primarily by an incorrect determination of the position of peak intensity, $\lambda_0$. All of the BR asymmetries quoted in this paper use velocity intervals of $\pm[50, 150]$ km s$^{-1}$ for the line wings and $[-30, 30]$ km s$^{-1}$ for the line core.

---

[1] The spline fit integrations are performed on a wavelength subgrid that is 99 times finer than the original data. The factor must be an odd number for technical reasons.

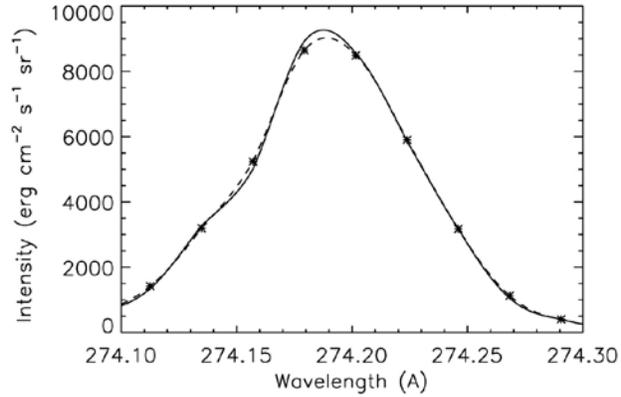 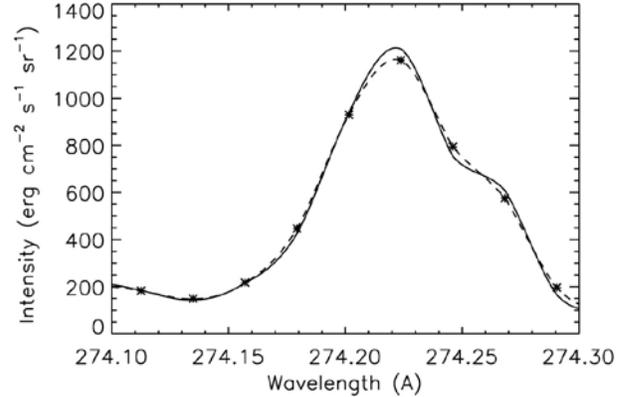

**Figure 5.** Line profile from raster position A in Figure 1. Stars--observed intensities; dashed curve--traditional spline fit; solid curve--spline fit with ICSI.

**Figure 6.** Line profile from raster position C in Figure 1. Stars--observed intensities; dashed curve--traditional spline fit; solid curve--spline fit with ICSI.

    Figure 5 shows a second example, from raster position (37,140) (position A). This time the solution takes 10 iterations to converge. The BR asymmetry is -0.8% (red excess) with ICSI and 1.2% (blue excess) without. Figure 6 shows a final example, from position (196,150) (position C). Convergence requires 9 iterations, and the asymmetries are 7.4% and 7.3% with and without ICSI, respectively.

    Figure 7 show the distribution of BR asymmetry errors across the entire 256x256 arcsec$^2$ raster. The error is defined to be the difference between the value determined with the traditional fit and the ICSI value. The errors are typically a few percent and are largely independent of the magnitude of the asymmetry itself. Thus, they can be a sizable fraction of the asymmetry where the asymmetry is small (most of the active region and quiet Sun), but are only a small fraction of the asymmetry where the asymmetry is large (faint "outflow regions" at the periphery of the active region). The distribution is skewed slightly toward negative errors, meaning that the ICSI value is more often larger (more positive or less negative) than the traditional fit, implying relatively more blue wing emission. Again this is due mostly to the improved determination of $\lambda_0$.

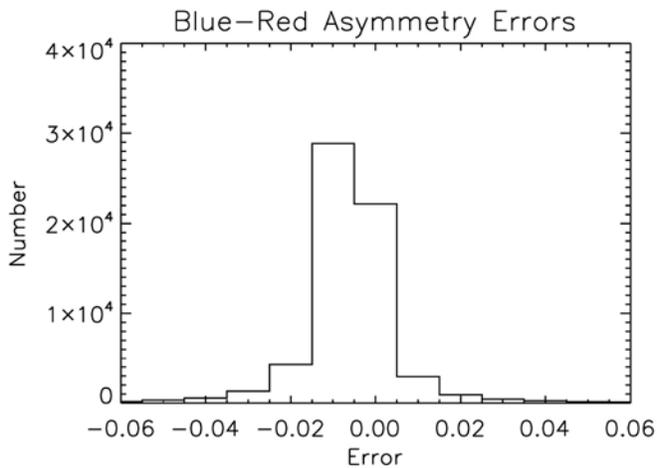

**Figure 7.** Distribution of BR asymmetry errors for the raster in Figure 1. The error is the difference between the value determined with the traditional fit and the ICSI value.

    Notice the localized enhancements, or bumps, in the blue wing of Figure 5 and the red wing of Figure 6. Such features are not uncommon in spectral lines and can be due to a distinct flow component or to a line blend. Determining the properties of these

7**Table 1**
Fit Parameter Errors

| Parameter | Main Component | Red Component |
|---|---|---|
| Central Intensity (%) | 4.9 | -10.6 |
| Width (%) | 8.5 | 3.4 |
| Doppler shift (km s$^{-1}$) | 1.1 | 1.1 |

features (intensity, velocity, width) is often important. ICSI can be very helpful in reducing the errors. We have performed a double Gaussian plus linear background fit to the profile in Figure 6, using both the original data and data corrected with ICSI. Corrected intensities have been adjusted up or down to match the ICSI solution at the wavelength bin centers. Table 1 gives the errors in the central intensity, width, and Doppler shift for both the main and red components of the double Gaussian fit (the difference in fit parameters using original and corrected data). The errors approach and even exceed 10%, which is substantial.

## 3. DISCUSSION

Spectral data consist of intensity values assigned to discrete wavelength positions. These values represent the average intensities over finite wavelength bins, and not the discrete intensities at the bin centers. In general, the two are different due to the curvature of the profile. Standard fitting routines (spline, Gaussian, etc.) do not account for this difference and attempt to find a continuous function that passes directly though the data points. As a consequence, the fits do not conserve intensity; within each bin, the area under the fit does not equal the observed intensity.

The iterative scheme given by Equations (1)-(5) corrects for this effect and provides a fit that conserves intensity. The scheme can be used with any fitting function (single or multiple Gaussian, kappa distribution, etc.). We choose a cubic spline for the work here because we wish to study line profiles of unknown shape that result from unresolved flows. We have developed a computationally fast and easy to use analysis routine called Intensity Conserving Spline Interpolation. It reduces errors in measurements of BR asymmetries that can be a sizable fraction of the asymmetries themselves.

In the course of preparing our manuscript, we were informed of other spline interpolation techniques that conserve intensity (Delhez 2003; Lang and Xu 2012). As far as we are aware, no routines based on these techniques are publicly available. A version of ICSI that runs under the IDL programming language has been submitted to the SolarSoft library and can be obtained from the authors upon request.

As emphasized, the iterative scheme that is at the heart of ICSI can be used with any fitting function. The ICSI code can be modified to call a different fitting routine from the cubic spline routine that is currently called. A simpler alternative is to preprocess the data with ICSI and then apply whatever fitting routine is desired. This was done in the example of a double Gaussian fit in the previous section. Given the rapid convergence, speed of computation, and ease of use, we suggest that ICSI be made a standard component of the processing pipeline for all spectroscopic data.

We close by noting that the iterative scheme behind ICSI can be used with any data in which the measured values represent averages over finite intervals. Examples include images with finite pixels and time series with finite temporal resolution.

This work of J.A.K. was supported by the NASA Supporting Research and Technology Program. The work of S.P. was supported by an FP7 Marie Curie Grant (FP7-PEOPLE-2010-RG/268288) and jointly by the European Union and Greek national government through the Operational Program "Education and Lifelong Learning" of the National Strategic Reference Framework (NSRF) Research Funding Program "Thales: Investing in knowledge society through the European Social Fund." D.T. acknowledges the Max-Planck partner Group of MPS at IUCAA. The authors benefited from participation in the International Space Science Institute team on Using Observables to Settle the Question of Steady vs. Impulsive Coronal Heating, led by Stephen Bradshaw and Helen Mason.

# REFERENCES

Bradshaw, S. J., & Cargill, P. J. 2013, ApJ, 770, 12
Bradshaw, S. J., & Klimchuk, J. A.  2015, ApJ, submitted
Brooks, D. H., & Warren, H. P.  2012, ApJ Lett, L5
Bryans, P., Young, P. R., & Doschek, G. A.  2010, ApJ, 715, 1012
Culhane, J. L., Harra, L. K., James, A. M., et al.  2007, SoPh, 243, 19
Delhez, E. J. M.  2003, Applied Math. Letters, 16, 17
De Pontieu, B., McIntosh, S. W., Hansteen, V. H., & Schrijver, C.  2009, ApJ Lett, 701, L1
De Pontieu, B., et al.  2011, Science, 331, 55
Doschek, G. A.  2012, ApJ, 754, 153
Hara, H., Watanabe, T., Harra, L. K., et al.  2008, ApJ Lett, 678, L67
Kitagawa, N., & Yokoyama, T.  2015, ApJ, 805, 97
Klimchuk, J. A.  2006, SoPh, 234, 41
Klimchuk, J. A.  2012, JGR, 117, A12102, doi:10.1029/2012JA018170
Klimchuk, J. A.  2015, Phil. Trans. R. Soc. A, 373: 20140256, doi:10.1098/rsta.2014.0256
Klimchuk, J. A., & Bradshaw, S. J.  2014, ApJ, 791, 60
Lang, F.-G., Xu, X.-P.  2012, J. Comp. Appl. Math., 236, 4214
Martinez-Sykora, J., De Pontieu, B., Hansteen, V., & McIntosh, S. W.  2011, ApJ, 732, 84
McIntosh, S. W., & De Pontieu, B.  2009, ApJ, 707, 524
Patsourakos, S., & Klimchuk, J. A.  2006, ApJ, 647, 1452
Patsourakos, S., Klimchuk, J. A., & Young, P. R.  2014, ApJ, 781, 58
Reale, R.  2010, Living Rev. Solar Phys. 7, 5
Tian, H., McIntosh, S. W., De Pontieu, B., et al.  2011, ApJ, 738, 18
Tian, H., McIntosh, S. W., Xia, L., et al.  2012, ApJ, 748, 106
Tripathi, D., & Klimchuk, J. A.  2013, ApJ, 779, 1